\documentclass[a4paper]{jpconf}
\usepackage[dvipdfmx]{graphicx}

\def\lesssim{\ \raise.3ex\hbox{$<$}\kern-0.8em\lower.7ex\hbox{$\sim$}\ }
\def\gesim{\ \raise.3ex\hbox{$>$}\kern-0.8em\lower.7ex\hbox{$\sim$}\ }

\graphicspath{{./epsfile/}}
\usepackage{amsmath}% required for `\cases' (yatex added)
\begin{document}
\title{Pseudogap phenomena in a two-dimensional ultracold Fermi gas near the Berezinskii-Kosterlitz-Thouless transition}
\author{M. Matsumoto, and Y. Ohashi}
\address{Department of Physics, Keio University, 3-14-1 Hiyoshi,
Kohoku-ku, Yokohama, Kanagawa 223-8522, Japan}
\ead{moriom@rk.phys.keio.ac.jp}
\begin{abstract}
 We investigate single-particle excitations and strong-coupling effects in a two-dimensional Fermi gas. Including pairing fluctuations within a Gaussian fluctuation theory, we calculate the density of states $\rho(\omega)$ near the Berezinskii-Kosterlitz-Thouless (BKT) transition temperature $T_{\rm BKT}$. Near $T_{\rm BKT}$, we show that superfluid fluctuations induce a pseudogap in $\rho(\omega)$. The pseudogap structure is very similar to the BCS superfluid density of states, although the superfluid order parameter is absent in the present two-dimensional case. Since a two-dimensional $^{40}$K Fermi gas has recently been realized, our results would contribute to the understanding of single-particle properties near the BKT instability. 
\end{abstract}
%%%%%%%%%%%%%%%%%%%%%%%%%%%%%%%%%%%%%%%%%%%%%%%%%%%%%%%%%%%%%%%%%%%%%%%%%%%%%%
\par
\section{Introduction}
\par
In ultracold Fermi gases, we can systematically study various many-body phenomena\cite{Gurarie,Bloch}. For example, a tunable pairing interaction associated with a Feshbach resonance enables us to examine superfluid properties from the weak-coupling regime to the strong-coupling limit\cite{Giorgini}. Another example is that, using an optical lattice, we can tune the dimensionality of the system\cite{Sommer}. Using the former advantage, the BCS-BEC crossover has been realized in $^{40}$K\cite{Jin} and $^6$Li\cite{Zwierlein,Kinast,Bartenstein} Fermi gases. Using the latter, a two-dimensional Fermi gas has recently been realized\cite{Sommer,Kohl,Frohlich}.
\par
In contrast to a three-dimensional Fermi gas, the BCS-type superfluid
phase transition is prohibited by strong pairing fluctuations in the
two-dimensional case\cite{Mermin,Hohenberg}. However,
Berezinskii\cite{Berezinskii1,Berezinskii2}, Kosterlitz and Thouless (BKT)\cite{Kosterlitz} clarified that a two-dimensional system may have a quasi-long-range order, exhibiting superfluidity. In cold atom physics, the BKT transition has recently been realized in a $^{87}$Rb Bose gas, loaded on a two-dimensional optical lattice\cite{Dalibard}.  The BKT transition in a Fermi gas is a crucial next challenge.
\par
Since the BKT transition occurs under the situation that the BCS long-range-order is completely suppressed by strong two-dimensional pairing fluctuations, physical properties near the BKT instability would be also affected by these fluctuations. In addition, in the three-dimensional BCS state, the superfluid order parameter $\Delta$ is directly related to the single-particle excitation gap. Since the pair formation also occurs in the fermionic BKT state, it is an interesting problem how the energy gap in the BKT phase is described without the gap parameter $\Delta$.
\par
In this paper, we investigate the single-particle density of states
$\rho(\omega)$ in a two-dimensional Fermi gas. Within a Gaussian
fluctuation theory\cite{NSR,Randeria}, we examine how pairing
fluctuations affect this quantity. In the three-dimensional case, they
are known to cause the pseudogap phenomenon\cite{Tsuchiya}, which is
characterized by a dip structure in $\rho(\omega)$ around $\omega=0$. We
show that this many-body phenomenon also occurs in the two-dimensional
case, where the pseudogapped density of states near $T_{\rm BKT}$ is
very similar to the BCS superfluid density of states. We briefly note
that pseudogap phenomena enhanced by the low-dimensionality of the
system have recently been discussed in Refs. \cite{Pietila,Watanabe,Klimin}.
\par
In this paper, we set $\hbar=k_{\rm B}=1$, and the system area is taken to be unity, for simplicity.
\par
%%%%%%%%%%%%%%%%%%%%%%%%%%%%%%%%%%%%%%%%%%%%%%%%%%%%%%%%%%%%%%%%%%%%%%%%%%%%%%
\par
\section{Formulation}
\par
We consider a model two-dimensional Fermi gas, consisting of two atomic
hyperfine states described by pseudospin
$\sigma=\uparrow,\downarrow$. We evaluate the BKT phase transition
temperature by the method used in Refs. \cite{Iskin,Tempere,Salanish}, which is based on the functional integral formalism\cite{Randeria} for the partition function $\mathcal{Z}=\int {\mathcal D}[\Psi,\bar{\Psi}]e^{-S[\Psi,\bar{\Psi}]}$ (which is related to the thermodynamic potential $\Omega$ as $\Omega=-T\ln{\mathcal Z}$). The action $S[\Psi,\bar{\Psi}]$ is given by
\begin{equation}
 S[\Psi,\bar{\Psi}]=\int_{0}^{\beta}d\tau \int d\mathbf{r}\biggr[\sum_{\sigma}\bar{\Psi}_{\sigma}(x)
[\partial_\tau+\xi_{\hat {\mathbf p}}]
 \Psi_{\sigma}(x)-U \bar{\Psi}_{\uparrow}(x)\bar{\Psi}_{\downarrow}(x)\Psi_{\downarrow}(x)\Psi_{\uparrow}(x)  \biggl],
\label{eq2}
\end{equation}
where $\beta=1/T$, and $\tau$ is the imaginary time. Grassmann variables $\Psi_\sigma(x)$ and ${\bar \Psi_\sigma}(x)$ describe Fermi atoms with an atomic mass $m$, where $x=(\mathbf{r},\tau)$. $\xi_{\hat {\mathbf p}}=-\nabla^2/(2m)-\mu$, where $\mu$ is the chemical potential. $-U(<0)$ is a pairing interaction, which is related to the $s$-wave scattering length $a_{\rm{s}}$ as\cite{Burnett},
$
-1/U=(m/2\pi)\ln(2/Ck_{\rm F}a_{\rm s})-\sum_{p\ge k_{\rm F}/\sqrt{2}}m/p^2,
$
where $k_{\rm F}$ is the Fermi momentum, and $C\simeq 1.78$ is the Euler constant.
\par
We introduce the Cooper-pair field $\Delta(x)$ through the Hubbard-Stratonovich transformation\cite{Randeria}. Carrying out the fermion integrals, we obtain
\begin{equation}
{\mathcal Z}=\int \mathcal{D}[\Delta,\Delta^{\ast}]e^{-S_{\rm{eff}}},
\label{eqZ}
\end{equation}
where
\begin{equation}
S_{\rm{eff}}[\Delta,\Delta^{\ast}]=\int_{0}^{\beta}\int d\mathbf{r}\frac{|\Delta(x)|^2}{U}-{\rm{Tr}}\ln[-\mathbf{G}^{-1}],
\label{eq8}
\end{equation}
and the Nambu-Gorkov Green's function $\mathbf{G}$ is given by
\begin{eqnarray}
\mathbf{G}^{-1}(x,x^{\prime})=
\left(
\begin{array}{cc}
-\partial_{\tau}-\xi_{\hat {\mathbf{p}}}&\Delta(x)\\
\Delta^{\ast}(x)&-\partial_{\tau}+\xi_{\hat {\mathbf{p}}}
\end{array}
\right)\delta(x-x^{\prime}).
\label{eq9}
\end{eqnarray}
In the normal state, we include pairing fluctuations described by $\Delta(x)$ within the Gaussian fluctuation level\cite{Randeria}. Executing the functional integrals in terms of $\Delta(x)$ and $\Delta^*(x)$ in Eq. (\ref{eqZ}), we obtain
$\Omega=\Omega_0
+T\sum_{\mathbf{q},i\nu_{n}}\ln[{1-U\Pi(\mathbf{q},i\nu_{n})]}
$. Here, $\Omega_0$ is the thermodynamic potential for a two-dimensional free Fermi gas, and
\begin{equation}
\Pi(\mathbf{q},i\nu_{n})=\sum_{\mathbf{p}}\frac{1-f(\xi_{\mathbf{p+\frac{q}{2}}})-f(\xi_{\mathbf{-p+\frac{q}{2}}})}{\xi_{\mathbf{p+\frac{q}{2}}}+\xi_{\mathbf{-p+\frac{q}{2}}}-i\nu_n}
\label{eq14}
\end{equation}
is the pair-correlation function (where $\xi_{\mathbf p}=p^2/(2m)-\mu$, $f(\varepsilon)=1/[e^{\beta\varepsilon}+1]$, and $\nu_n$ is the boson Matsubara frequency). The equation for the number $N$ of Fermi atoms is then obtained from $\Omega$ as
\begin{equation}
N=-{\partial \Omega \over \partial\mu}=2T\sum_{{\mathbf p},i\omega_n}{\tilde G}({\mathbf p},i\omega_n).
\label{eq14a}
\end{equation}
In Eq. (\ref{eq14a}), the single-particle thermal Green's function ${\tilde G}$ has the form
\begin{equation}
{\tilde G}({\mathbf p},i\omega_n)
=G_0({\mathbf p},i\omega_n)
+G_0({\mathbf p},i\omega_n)\Sigma({\mathbf p},i\omega_n)
G_0({\mathbf p},i\omega_n),
\label{eq14b}
\end{equation}
where $G_0({\mathbf p},i\omega_n)=1/[i\omega_n-\xi_{\mathbf p}]$ is the bare Green's function, $\omega_n$ is the fermion Matsubara frequency, and the self-energy
\begin{equation}
\Sigma({\mathbf p},i\omega_n)=T\sum_{\mathbf{q},i\nu_{n}}\Gamma(\mathbf{q},i\nu_{n})G_{0}(\mathbf{q-p},i\nu_{n}-i\omega_{n})
\label{eq14c}
\end{equation}
is just the same as that in the $T$-matrix approximation\cite{Tsuchiya}. $\Gamma(\mathbf{q},i\nu_{n})=-U/[1-U\Pi(\mathbf{q},i\nu_{n})]$ in Eq. (\ref{eq14c}) is the particle-particle scattering matrix. 
\par
We determine the chemical potential $\mu$ from Eq. (\ref{eq14a}). We then calculate the density of states $\rho(\omega)$ from the Green's function $G^{-1}=G_0^{-1}-\Sigma$ with the self-energy in Eq. (\ref{eq14c}) as\cite{note},
\begin{equation}
\rho(\omega)=-\frac{2}{\pi}\sum_{\mathbf{p}}{\rm{Im}}\bigl[G(\mathbf{p},i\omega_{n}\rightarrow \omega+i\delta)\bigr].
\label{eq17}
\end{equation}
\par
%%%%%%%%%%%%%%%%%%%%%%%%%%%%%%%%%%%%%%%%%%%%%%%%%%%%%%%%%%%%%%%%%%%%%%%%%%%%%%%
\begin{figure}[t]
\begin{center}
\includegraphics[width=0.715\linewidth,keepaspectratio]{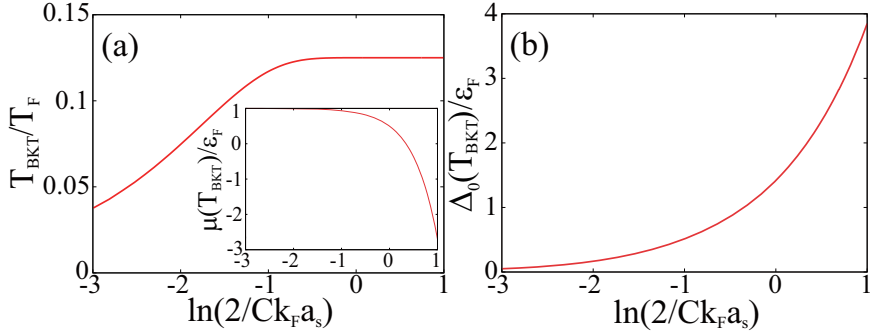}
\end{center}
\caption{(a) Calculated BKT phase transition temperature $T_{\rm{BKT}}$. The inset shows the chemical potential $\mu(T_{\rm{BKT}})$. $T_{\rm F}$ is the Fermi temperature. (b) Saddle point solution (mean-field order parameter) $\Delta_{0}$ at $T_{\rm{BKT}}$, normalized by the Fermi energy $\varepsilon_{\rm F}$.}
\label{fig1}
\end{figure}
%%%%%%%%%%%%%%%%%%%%%%%%%%%%%%%%%%%%%%%%%%%%%%%%%%%%%%%%%%%%%%%%%%%%%%%%%%%%%%%
\par
While the present formalism can describe strong-coupling effects in the
normal state, it does not give the BKT instability. Thus, to determine
$T_{\rm BKT}$, we employ the prescription used in Refs. \cite{Iskin,Tempere,Salanish}. Including phase fluctuations $\theta(x)$ around the saddle point solution $\Delta_0$ of the action $S_{\rm{eff}}$ in Eq. (\ref{eq8}) ($\Delta(x)=\Delta_0e^{i\theta(x)}$) within the Gaussian fluctuations level, we obtain
\begin{equation}
\mathcal{Z}=e^{-S_{\rm{MF}}}\int \mathcal{D}[\theta]e^{-S_{\rm{FL}}},
\label{eq20}
\end{equation}
where $S_{\rm {MF}}$ is given by Eq. (\ref{eq8}) with $\Delta(x)$ being replaced by $\Delta_0$. 
$S_{\rm{FL}}=\sum_{\mathbf{q},i\nu_{n}}[Jq^2+K\nu_{n}^2]\theta(\mathbf{q},i\nu_{n})\theta(-\mathbf{q},-i\nu_{n})$ is the fluctuation contribution. Here,
\begin{equation}
   J=\frac{1}{4m}\sum_{\mathbf{p}}\biggr[1-\frac{\xi_{\mathbf{p}}}{E_{\mathbf{p}}}{\rm{tanh}}\biggl(\frac{E_{\mathbf{p}}}{2T}\biggr)-\frac{1}{2T}\frac{p^2}{2m}{\rm sech}^{2}\biggl({E_{\mathbf p} \over 2T}\biggr)
    \biggl]
\label{eq22}
 \end{equation}
is the phase stiffness, and
$
   K=\frac{1}{4}\sum_{\mathbf{p}}\biggr[\frac{\Delta_{0}^2}{E_{\mathbf{p}}^3}{\rm{tanh}}\bigl(\frac{E_{\mathbf{p}}}{2T}\bigr)
+\frac{1}{2T}\frac{\xi^{2}_{\mathbf{p}}}{E_{\mathbf{p}}^{2}}
{\rm sech}^{2}\bigl(\frac{E_{\mathbf{p}}}{2T}\bigr)
    \biggr],
$
 where $E_{\mathbf{p}}=\sqrt{\xi^2_{\mathbf{p}}+\Delta^2_{0}}$. Using the phase stiffness constant $J$ in Eq. (\ref{eq22}), we determine $T_{\rm BKT}$ from the KT-Nelson formula\cite{Nelson}, $T_{\rm BKT}=(\pi/2)J(T_{\rm BKT})$. In this equation, $\Delta_0$ is determined from the BCS gap equation,
$
1=U\sum_{\mathbf p}{1 \over 2E_{\mathbf p}}
\tanh \bigl({E_{\mathbf p} \over 2T}\bigr),
$
and $\mu$ is evaluated from the number equation $N=-\partial \Omega/\partial \mu$, where $\Omega=-T\ln{\mathcal Z}$ with ${\mathcal Z}$ being given by Eq. (\ref{eq20}) where the $\theta$-integration has been executed. Figure \ref{fig1} shows the self-consistent solutions for $T_{\rm BKT}$, $\mu$, and $\Delta_0$. 
\par
%%%%%%%%%%%%%%%%%%%%%%%%%%%%%%%%%%%%%%%%%%%%%%%%%%%%%%%%%%%%%%%%%%%%%%%%%%%%%%%
\begin{figure}[t]
\begin{center}
\includegraphics[width=0.8\linewidth,keepaspectratio]{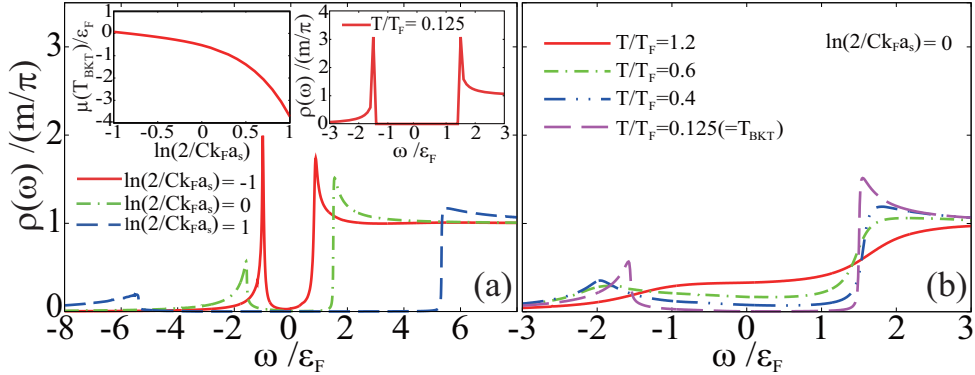}
\end{center}
\caption{(a) Calculated density of states $\rho(\omega)$ at $T_{\rm BKT}$. The left inset shows the chemical potential $\mu$ used in this calculation. The right inset shows the BCS superfluid density of states, where the parameter set $(\Delta_0,\mu)$ shown in Fig. \ref{fig1} at $\ln(2/Ck_{\rm F}a_{\rm s})=0$ is used. (b) $\rho(\omega)$ at various temperatures when $\ln(2/Ck_{\rm F}a_{\rm s})=0$.}
\label{fig2}
\end{figure}
%%%%%%%%%%%%%%%%%%%%%%%%%%%%%%%%%%%%%%%%%%%%%%%%%%%%%%%%%%%%%%%%%%%%%%%%%%%%%%%
\par
\section{Pseudogap phenomenon near the BKT phase transition} 
\par
Figure \ref{fig2}(a) shows the single-particle density of states $\rho(\omega)$ at $T_{\rm BKT}$. We see a clear gap-like structure even in the relatively weak-coupling case ($\ln(2/Ck_{\rm F}a_{\rm s})=-1$), which becomes more remarkable for a stronger interaction. We recall that, although we have introduced the saddle point solution $\Delta_0$ in calculating $T_{\rm BKT}$, this mean-field order parameter is {\it not} used in $\rho(\omega)$ in Eq. (\ref{eq17}). Thus, the gap-like structure in Fig. \ref{fig2}(a) is the pseudogap associated with pairing fluctuations. Since the low-dimensionality enhances pairing fluctuations, the pseudogap in the present case is more remarkable than the three-dimensional case\cite{Tsuchiya}.
\par
When $\ln(2/Ck_{\rm F}a_{\rm s})\ge 0$, the pseudogap structure at $T_{\rm BKT}$ is very similar to the BCS superfluid density of states with sharp coherence peaks at the gap edges (see the right inset in Fig. \ref{fig2}(a)). To understand this, we note that the particle-particle scattering matrix $\Gamma({\mathbf q},i\nu_n)$ in Eq. (\ref{eq14c}), which describes fluctuations in the Cooper channel, would be large when ${\mathbf q}=\nu_n=0$ near $T_{\rm BKT}$. Using this, when we approximate Eq. (\ref{eq14c}) to $\Sigma({\mathbf p},i\omega_n)\simeq G_{0}(\mathbf{-p},-i\omega_{n})T\sum_{\mathbf{q},i\nu_{n}}\Gamma({\mathbf q},i\nu_n)\equiv -G_{0}(\mathbf{-p},-i\omega_{n})\Delta_{\rm PG}^2$, the Green's function $G$ in Eq. (\ref{eq17}) is found to have the same form as the ordinary particle component of the BCS Green's function as,
\begin{equation}
 G(\mathbf{p},i\omega_{n})=-\frac{i\omega_{n}+{\xi}_{\mathbf{p}}}{\omega_{n}^2+\xi_{\mathbf{p}}^2+\Delta_{\rm{PG}}^2}.
 \label{eq18}
\end{equation}
This means that the so-called pseudogap parameter $\Delta_{\rm PG}$\cite{Tsuchiya,Levin} plays similar roles to the BCS superfluid order parameter $\Delta_0$, leading to the pseudogap with sharp `coherence' peaks, as shown in Fig. \ref{fig2}(a). Thus, when one measures $\rho(\omega)$ near $T_{\rm BKT}$, the system would look like the mean-field BCS state, although the superfluid order parameter $\Delta$ is actually absent and strong pairing fluctuations only exist.
\par
While the pseudogap structure is similar to the BCS density of states at $T_{\rm BKT}$, their temperature dependence is very different from each other. In the BCS state, when the temperature increases, the gap width gradually becomes small, keeping the sharp coherence peaks at the gap edges, In contrast, Fig. \ref{fig2}(b) shows that the pseudogap structure, as well as the sharp peaks, gradually become obscure with increasing the temperature. 
\par
%%%%%%%%%%%%%%%%%%%%%%%%%%%%%%%%%%%%%%%%%%%%%%%%%%%%%%%%%%%%%%%%%%%%%%%%%%%%%%
\par
\section{Summary}
\par
To summarize, we have discussed strong-coupling properties of a two-dimensional ultracold Fermi gas. Including pairing fluctuations within the framework of a Gaussian fluctuation theory, we calculated the single-particle density of states $\rho(\omega)$ near the BKT phase transition. At $T_{\rm BKT}$, pairing fluctuations induce a pseudogap in $\rho(\omega)$, the structure of which is very similar to the BCS superfluid density of states with a finite excitation gap, as well as sharp coherence peaks at the gap edges. That is, the BCS-like energy gap is expected to be observed even in a two-dimensional Fermi gas near $T_{\rm BKT}$, although the superfluid order parameter is absent and the system is dominated by low-dimensional pairing fluctuations. Since the achievement of the BKT phase transition is an exciting challenge in cold Fermi gas physics, our results would be useful for the study of this superfluid state on the view point of single-particle excitations.
\par
%%%%%%%%%%%%%%%%%%%%%%%%%%%%%%%%%%%%%%%%%%%%%%%%%%%%%%%%%%%%%%%%%%%%%%%%%%%%%
\ack
We thank D. Inotani, R. Hanai, and H. Tajima for discussions. Y.O. was supported by a Grant-in-Aid for Scientific Research from MEXT in Japan (Grant No. 25105511 and No. 25400418).
\par
%%%%%%%%%%%%%%%%%%%%%%%%%%%%%%%%%%%%%%%%%%%%%%%%%%%%%%%%%%%%%%%%%%%%%%%%%%%%%
\par


\section{References}
\medskip
\begin{thebibliography}{99}
\bibitem{Gurarie} Gurarie V and Radzihovsky L 2007 {\it Ann. Phys.} {\bf 332} 2 
\bibitem{Bloch} Bloch I, Dalibard J and Zwerger W 2008 {\it Rev. Mod. Phys.} {\bf 80} 885
\bibitem{Giorgini} Giorgini S, Pitaevskii L P and Stringari S 2008 {\it Rev. Mod. Phys.} \textbf{80} 1215
\bibitem{Sommer} Sommer A T, Cheuk L W, Ku M J H, Bakr W S and Zwierlein M W 2012 {\it Phys. Rev. Lett.} \textbf{108} 045302
\bibitem{Jin} Regal C A, Greiner M and Jin D S 2004 {\it Phys. Rev. Lett.} \textbf{92} 040403
\bibitem{Zwierlein} Zwierlein M W, Stan C A, Schunck C H, Raupach S M F, Kerman A J and Ketterle W 2004 {\it Phys. Rev. Lett.} \textbf{92} 120403
\bibitem{Kinast} Kinast J, Hemmer S L, Gehm M E, Turlapov A and Thomas J E 2004 {\it Phys. Rev. Lett.} \textbf{92} 150402
\bibitem{Bartenstein} Bartenstein M, Altmeyer A, Riedl S, Jochim S, Chin C, Denschlag J H and Grimm R 2004 {\it Phys. Rev. Lett.} \textbf{92} 203201
\bibitem{Kohl} Feld M, Fr\"{o}hlich B, Vogt E, Koschorreck M and K\"{o}hl M 2011 \textit{Nature} \textbf{480} 75
\bibitem{Frohlich} Fr\"{o}hlich B, Feld M, Vogt E, Koschorreck M, Zwerger W and K\"{o}hl M 2011 {\it Phys. Rev. Lett.} \textbf{106} 105301
\bibitem{Mermin} Mermin N D and Wagner H 1966 {\it Phys. Rev. Lett.} {\bf 17} 1133
\bibitem{Hohenberg} Hohenberg P C 1967  {\it Phys. Rev.} {\bf 158} 383
\bibitem{Berezinskii1} Berezinskii V L 1971 {\it Sov. Phys. JETP} {\bf 32} 493
\bibitem{Berezinskii2} Berezinskii V L 1972 {\it Sov. Phys. JETP} {\bf 34} 610
\bibitem{Kosterlitz} Kosterlitz J M and Thouless D J 1973 {\it
	J. Phys. C:Solid State Phys.} {\bf 6} 1181 
\bibitem{Dalibard} Hadzibabic Z, Kr$\ddot{\rm{u}}$ger P, Cheneau M, Battelier B and Dalibard J 2006 {\it Nature} {\bf 441} 1118
\bibitem{NSR} Nozi\`eres P and Schmitt-Rink S 1985 {\it J. Low Temp. Phys.} {\bf 59} 195
\bibitem{Randeria} S\'a de Melo C A R, Randeria M and Engelbrecht J R 1993 {\it Phys. Rev. Lett.} {\bf 71} 3202
\bibitem{Tsuchiya} Tsuchiya S, Watanabe R and Ohashi Y 2009 {\it Phys. Rev.} A {\bf 80} 033613
\bibitem{Pietila} Pietil\"a V 2012 {\it Phys. Rev.} A {\bf 86} 023608
\bibitem{Watanabe} Watanabe R, Tsuchiya S and Ohashi Y 2013 {\it Phys. Rev.} A {\bf 88} 013637
\bibitem{Klimin}  Klimin S N, Tempere J and Devreese J T 2012 {\it New J. Phys.} {\bf 14} 103044
\bibitem{Iskin} Iskin M and S\'a de Melo C A R 2009 {\it Phys. Rev. Lett.} {\bf 103} 165301
\bibitem{Tempere} Tempere J, Klimin S N and Devreese J T 2009 {\it Phys. Rev.} A {\bf 79} 053637
\bibitem{Salanish} Salasnich L, Marchetti P A and Toigo F 2013 {\it Phys. Rev.} A {\bf 88} 053612
\bibitem{Burnett} Morgan S A, Lee M D and Burnett K 2002 {\it Phys. Rev.} A {\bf 65} 022706
\bibitem{note} When the Green's function in Eq. (\ref{eq14b}) is used in Eq. (\ref{eq17}), the resulting density of states is known to unphysically become negative in the strong-coupling region\cite{Tsuchiya}. This difficulty is, however, avoided by using $G^{-1}=G_0^{-1}-\Sigma$. We note that the Green's function with the self-energy given by Eq. (\ref{eq14c}) is the same as that in the $T$-matrix approximation\cite{Tsuchiya}.
\bibitem{Nelson} Nelson D R and Kosterlitz J M 1977 {\it Phys. Rev. Lett.} {\bf 39} 1201
\bibitem{Levin} Chen Q J and Levin K 2009 {\it Phys. Rev. Lett.} {\bf 102} 190402
\end{thebibliography}
\end{document}